# On Natural Genetic Engineering:

# Structural Dynamism in Random Boolean Networks

Larry Bull

Department of Computer Science & Creative Technologies

University of the West of England

Bristol BS16 1QY, U.K.

+44 (0)117 3283161

Larry.Bull@uwe.ac.uk

Abstract

This short paper presents an abstract, tunable model of genomic structural change within the cell lifecycle and explores its use with simulated evolution. A well-known Boolean model of genetic regulatory networks is extended to include changes in node connectivity based upon the current cell state, e.g., via transposable elements. The underlying behaviour of the resulting dynamical networks is investigated before their evolvability is explored using a version of the *NK* model of fitness landscapes. Structural dynamism is found to be selected for in non-stationary environments and subsequently shown capable of providing a mechanism for evolutionary innovation when such reorganizations are inherited.

1. Introduction

Numerous mechanisms have been identified through which changes in DNA sequences can occur in ways other than copy errors, such as via transposable elements (e.g., see [Craig et al., 2002] for an overview). The significance of such mechanisms with respect to evolutionary innovation has recently been highlighted [Shapiro, 2011]. The gradual accumulation of random DNA copy errors does not appear to be the primary source of variation over which selection subsequently acts; instead, specific biochemical processes generate novelty through *context responsive* DNA restructuring during the cell or organismal history.

This paper begins by presenting an abstract model for consideration of such mechanisms within a simple genetic regulatory network (GRN). In particular, the model is an extension of a well-known Boolean GRN formalism. The paper then uses the presented generic scheme to examine how such restructuring processes affect evolvability using a recently presented extension to a well-known tunable fitness landscape model.

Random Boolean networks (*RBN*) [Kauffman, 1969] were introduced as an abstract model by which to explore aspects of genetic regulatory networks. *RBN* consist of $R$ genetic loci/nodes, each connected to $B$ other randomly chosen nodes, with each performing a randomly assigned Boolean update function based upon the current state of those nodes. The emergent behaviour of these discrete dynamical systems has been explored widely (e.g., see [Kauffman, 1993][Gershenson, 2002] for overviews). Some of this work has included the use of simulated evolution to induce specific network-wide behaviour (e.g., [Lemke et al., 2001]).

With the aim of enabling systematic exploration of the evolvability of such GRN, *RBN* have recently been combined with the *NK* model of fitness landscapes. The *NK* model [Kauffman & Levin, 1978] of fitness landscapes considers sets of $N$ (binary) genes or traits, each of which depends upon $K$ others within the set. As the degree of epistasis $K$ is varied, so features of the landscapes are affected. This abstract, tunable model has been used widely to explore aspects of evolution (e.g., see [Kauffman, 1993][Welch & Waxman, 2005] for overviews). In the combined form – termed the *RBNK* model [Bull, 2012] – a simple relationship between the states of $N$ randomly assigned nodes within an *RBN* was assumed such that their value is used within a given *NK* fitness landscape of trait dependencies. This paper adds dynamic connectivity based upon the current state of the

GRN during its lifecycle to the *RBN* and *RBNK* models to begin to explore the role of context dependent feedbacks on evolution.

## 2. Adding Structural Dynamism to the *RBN* Model

As noted above, within the traditional form of *RBN*, a network of *R* nodes, each with *B* directed connections randomly assigned from other nodes in the network, all update synchronously based upon the current state of those *B* nodes. Hence those *B* nodes are seen to have a regulatory effect upon the given node, specified by the given Boolean function arbitrarily attributed to it; the details of RNA and/or protein actions are abstracted out. Since they have a finite number of possible states and they are deterministic, such networks eventually fall into an attractor. It is well-established that the value of *B* affects the emergent behaviour of *RBN* wherein attractors typically contain an increasing number of states with increasing *B*. Three phases of behaviour were originally suggested through observation: ordered when *B*=1, with attractors consisting of one or a few states; chaotic when *B*≥3, with a very large number of states per attractor; and, a critical regime around *B*=2, where similar states lie on trajectories that tend to neither diverge nor converge (see [Kauffman, 1993] for discussions of this critical regime, e.g., with respect to perturbations). Subsequent formal analysis using an annealed approximation of behaviour also identified *B*=2 as the critical value of connectivity for behaviour change [Derrida & Pomeau, 1986]. Typical behaviour is shown in Figure 1 where the percentage of nodes changing state per update cycle indicates the size of the attractor reached.

Mobile genetic elements are DNA sequences that may be either copied or removed and then inserted at a new position in the genome [Shapiro, 2011]. For example, retrotransposons use an RNA copy of themselves, whereas DNA transposons rely upon specific proteins for their "cutting and pasting" into new sites. The targeting of a new position ranges from the very specific, typically by exploiting sequence recognition proteins, to more or less arbitrary movement. Transposons [McClintock, 1987] are found widely in both prokaryotes and eukaryotes, and they have been associated with many significant evolutionary innovations. For example, retrotransposons were involved in the genetic changes which separate humans from chimpanzees [Wang et al., 2005].

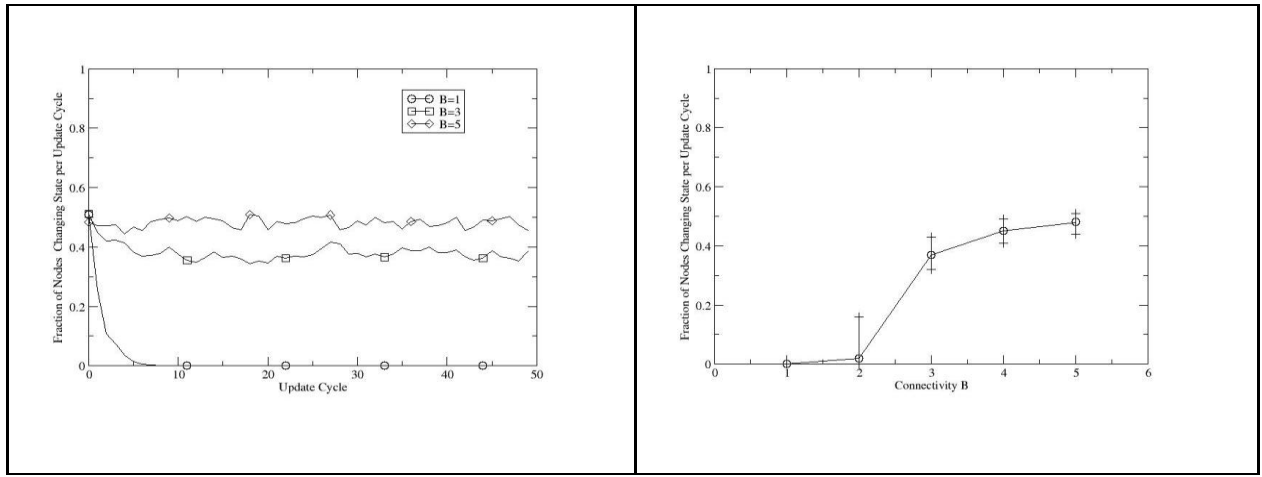

Figure 1: Typical behaviour of *RBN* with *R*=100 nodes: on the left, showing example temporal dynamics; and on the right, the average behaviour (100 runs) after 100 update cycles. Nodes were initialized arbitrarily. Error bars show the min and max behaviour.

To add structural dynamism through transposon-like alterations to the traditional *RBN* model, some fraction of the *R* nodes (analogous to genetic loci) are extended to include a second set of *B'* connections to randomly chosen nodes. Each such dynamic node also performs a randomly assigned rewiring function based upon the current state of the *B'* nodes, as shown in Figure 2. Hence on each cycle, each node updates its state based upon the current state of the *B* nodes it is connected to using the Boolean logic function assigned to it in the standard way. Then, if that node is also structurally dynamic, those *B* connections are altered according to the current state of the *B'* nodes it is connected to using its rewiring table. Hence the actions of the *B'* nodes may be seen as an abstraction of the production of a transposase enzyme(s), for example, which subsequently causes a restructuring in the transcription regulatory circuit, i.e., by moving the *B* connections for a given node.

Figure 3 shows the typical (dynamical) behaviour after 100 update cycles including varying percentages of rewiring nodes in *RBN* as described, for *R*=100 and various *B*. For simplicity, it is assumed *B*=*B'* throughout this paper. As can be seen, and as might be expected, for *B*>2 no significant effect is seen since such *RBN* are typically chaotic in the standard case. For *B*=1, the rewiring through the extra connectivity induces behaviour akin to *B*=2 in the traditional case, although the degree of variance in behaviour means the change is not significant (T-test, $p \geq 0.05$) for any percentage of rewiring nodes. For *B*=2 behaviour does not vary significantly from the traditional case when less than 50% of nodes are dynamic.

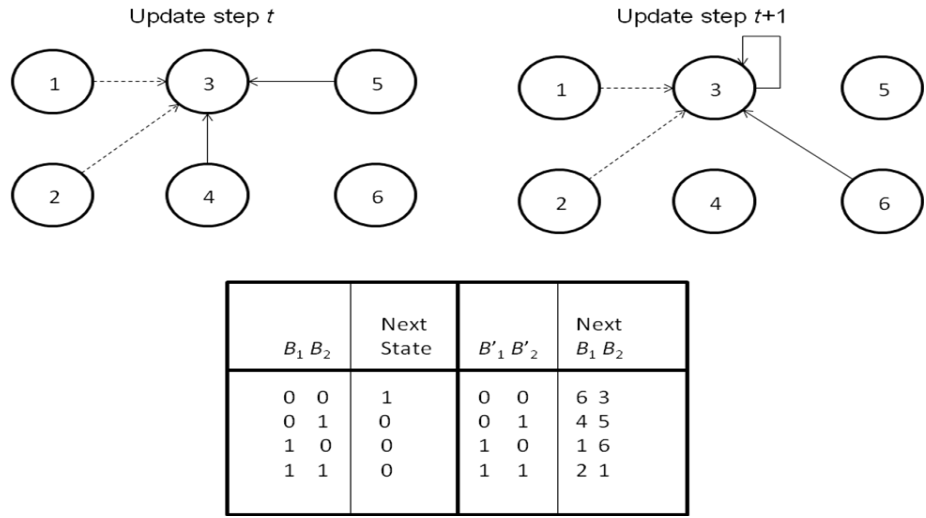

Figure 2: Example *RBN* with structural dynamism. The look-up table and connections for node 3 are shown in an *R*=6, *B*=2 network. Nodes capable of rewiring have *B'* extra structure regulation connections into the network (dashed arrows) and use the state of those nodes to alter the standard *B* transcription regulation connections (solid arrows) on the next update cycle (*B'*=2). Thus in the *RBN* shown, node 3 is a dynamic node and uses nodes 1 and 2 to determine any structural changes. At update step *t*, node 3 is shown using the states of nodes 4 and 5 to determine its state for the next cycle. Assuming both are at state '0', the given node above would transit to state '1' for the next cycle and source its *B* inputs from nodes 6 and 3 on that subsequent cycle, as defined in the first row of the table shown. A DNA transposon-like mediated change in the regulation network is said to have occurred and the genome rewritten – the *B* source connection ids are altered.

Hence the presence of dynamic nodes undertaking rewiring during the updating of *RBN* does appear to affect behaviour for low connectivity networks to such an extent they become chaotic. The next section considers the evolvability of such *RBN* using a variant of a well-known tuneable fitness landscape model which allows somewhat systematic exploration of behaviour.

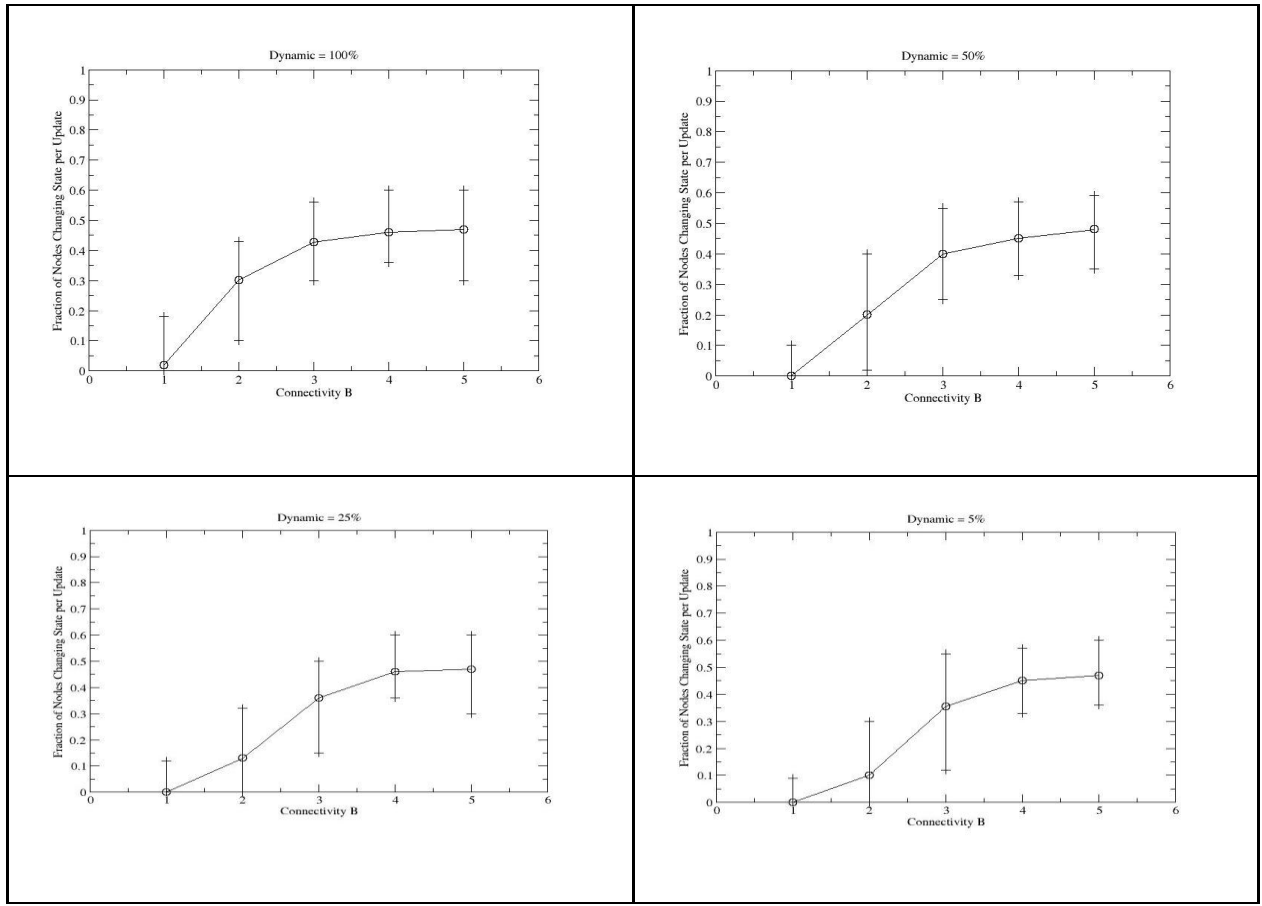

Figure 3: Typical behaviour (100 runs) of *RBN* with *R*=100 nodes and varying percentages of dynamic nodes after 100 update cycles. Initial node states and connections were assigned arbitrarily.

## 3. Evolving *RBN* with Structural Dynamism

### 3.1 Evolving *RBN*

There is a small amount of prior work exploring the use of simulated evolution to design *RBN*. Van den Broeck and Kawai [1990] explored the use of a simulated annealing-type approach to design feedforward *RBN* for the four-bit parity problem. Kauffman [1993] evolved *RBN* to match a given attractor. Lemke et al. [2001] did the same. The same approach has been used to explore attractor stability [Fretter et al., 2009] and to model real regulatory network data (e.g., [Tan & Tay, 2006]). Sipper and Ruppin [1997] evolved *RBN* for the well-known density task. Bull [2009] has evolved *RBN* ensembles to solve simple machine learning problems. See [Bull, 2012] for a more general overview.

3.2 The *RBNK* Model

As noted above, the *RBN* and *NK* models have recently been combined – the *RBNK* model – to explore GRN and phenotype dependency [Bull, 2012]. As shown in Figure 4, *N* nodes in the *RBN* are chosen as "outputs", i.e., their state determines fitness using the *NK* model. Kauffman and Levin [1987] introduced the *NK* model to allow the systematic study of various aspects of fitness optimization (see [Kauffman, 1993]). In the standard model an individual is represented by a set of *N* (binary) genes or traits, each of which depends upon its own value and that of *K* randomly chosen others in the individual. Thus increasing *K*, with respect to *N*, increases the epistatic linkage. This increases the ruggedness of the fitness landscapes by increasing the number of fitness peaks.

The *NK* model assumes all epistatic interactions are so complex that it is only appropriate to assign (uniform) random values to their effects on fitness. Therefore for each of the possible *K* interactions, a table of $2^{(K+1)}$ fitnesses is created, with all entries in the range 0.0 to 1.0, such that there is one fitness value for each combination of traits. The fitness contribution of each trait is found from its individual table. These fitnesses are then summed and normalised by *N* to give the selective fitness of the individual. Exhaustive search of *NK* landscapes [Smith & Smith, 1999] suggests three general classes exist: unimodal when *K*=0; uncorrelated, multi-peaked when *K*>3; and, a critical regime around 0<*K*<4, where multiple peaks are correlated. This differs slightly from Kauffman's [1993] analysis wherein it was suggested correlation decreases as $K \rightarrow N$. A difference perhaps caused by his use of non-exhaustive search to explore landscapes of increasing *N*. An element of correspondence between *NK* and *RBN* models with respect to the degree of connectivity and typical properties can therefore be noted.

The combination of the *RBN* and *NK* model enables a systematic exploration of the relationship between phenotypic traits and the genetic regulatory network by which they are produced. In this paper, following [Bull, 2012], a simple scheme is adopted: *N* phenotypic traits are attributed to arbitrarily chosen nodes within the network of *R* genetic loci but with environmental inputs applied to the first *N* loci (Figure 4). Hence the *NK* element creates a tuneable component to the overall fitness landscape with behaviour (potentially) influenced by the environment.

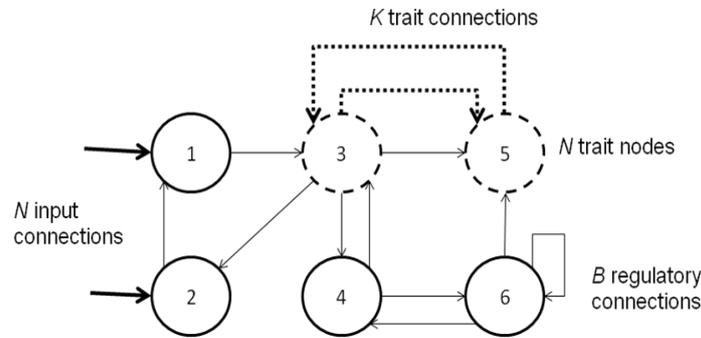

An $R$=6, $B$=2, $N$=2, $K$=1 network with two inputs

Figure 4: Variant of the *RBNK* model used with $N$ external inputs also applied thereby making the GRN sensitive to its environment. The $B'$ connections are not shown for clarity.

Following [Kauffman, 1993], the simple case of a greedy, genetic hillclimber is considered here. Each *RBN* is represented as a list of integers to define each node's start state, Boolean function, $B$ connection ids, $B'$ connection ids, connection changes table entries (see Figure 2), and whether it is a dynamic node or not. Mutation can therefore either (with equal probability): alter the Boolean function of a randomly chosen node; alter a randomly chosen $B$ connection (used as the initial connectivity if a dynamic node); alter a node start state; turn a node into or out of being a dynamic rewiring node; alter one of the rewiring entries in the look-up table if it is a dynamic node; or, alter a randomly chosen $B'$ connection, again only if it is a dynamic node. A single fitness evaluation of a given GRN is ascertained by updating each node for 100 cycles from the genome defined start states. At each update cycle, the value of each of the $N$ trait nodes in the GRN is used to calculate fitness on the given $NK$ landscape. The final fitness assigned to the GRN is the average over 100 such updates here. A mutated GRN becomes the parent for the next generation if its fitness is higher than that of the original. In the case of fitness ties the number of dynamic nodes is considered, with the smaller number favoured, with the decision being arbitrary upon a further tie. Hence there is a slight selective pressure against structural dynamism here.

## 4. Experimentation

4.1 Stationary and Non-stationary Fitness Landscapes

In the following, $R=100$, $N=10$ and results are the average of 100 runs (10 runs on each of 10 landscapes per parameter configuration). Evolution is run for 50,000 generations - when fitness stasis is typically seen. Nodes have a 50% probability of being dynamic upon initialization. As in [Bull, 2012], $0<B<5$ and $0≤K<5$ are used.

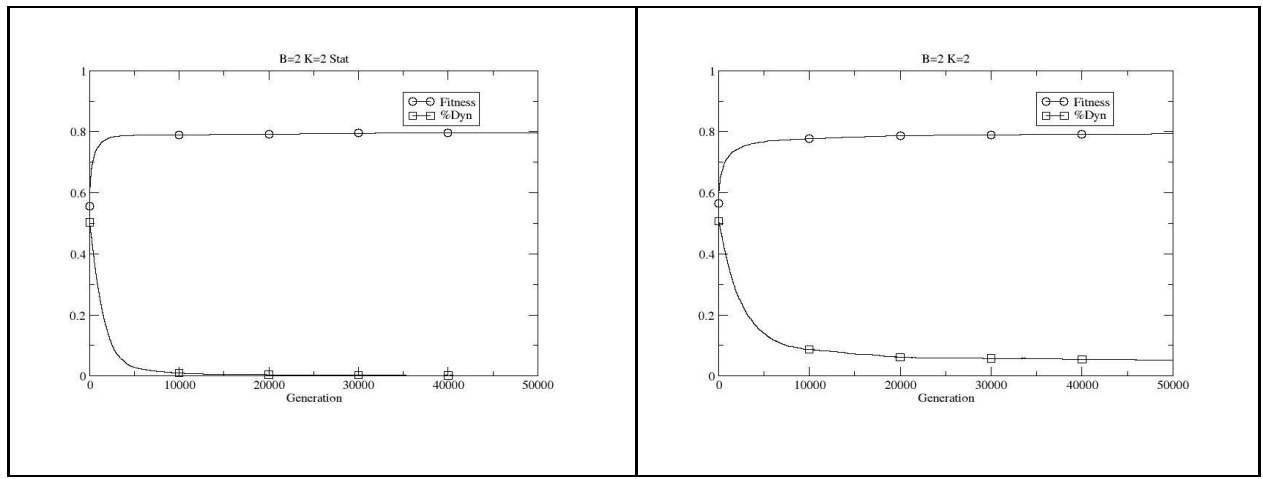

Figure 5: Examples of the dynamic *RBN* in a stationary (left) and non-stationary (right) environment.

Figure 5 shows examples of the typical evolutionary behavior of the structurally dynamic *RBN* on two types of *NK* landscape. On the left, a traditional single fitness landscape is used and a constant input of $N$ 0's applied to the *RBN*. As can be seen, the fraction of dynamic nodes within the GRN quickly decreases to zero. On the right, a non-stationary version of the model is used such that, after 50 update cycles on one *NK* landscape, with all 0's applied as the input, fitness is then ascertained from a second landscape for the remaining 50 cycles, with all 1's applied as the input. As can be seen, the percentage of dynamic nodes stabilizes at around 5%. Hence structural dynamism is selected for in the non-stationary case. This latter version of the model was somewhat motivated by the growing number of examples of environmentally triggered – typically under stress conditions - genomic rearrangements found in a wide variety of organisms (see [Shapiro, 2011]). Figure 6 shows further examples of how this result holds for all $B<3$, i.e., non-chaotic, GRN regardless of the underlying topology of the fitness landscapes (T-test, $p<0.05$). There is no significant difference in structure when $B>2$ (T-test, $p≥0.05$). Analysis of the rewiring behaviour in the low *B* cases shows that the dynamic nodes typically fire for *only* the first few update cycles after both initialization and the switch in input halfway through the lifecycle. Fitness is significantly decreased (T-test, $p<0.05$) for high *B* in all cases, as previously reported [Bull, 2012]. The same general result was found (not shown) for varying the size of the networks, e.g., $R=50$ or 200, with the final percentage of dynamic nodes tending to vary proportionally with *R*, i.e., ~2.5% and 10% respectively, but not with consistent statistical significance. Given that both the number of size of attractors are known to be proportional to *R* [Kauffman, 1993] this is perhaps to be expected. Thus it appears that *the evolutionary process is exploiting structural dynamism to help shape the attractor space of the RBN such that high fitness can be reliably reached depending upon the environmental input and GRN state*. The capacity for structural dynamism disappears in static environments.

Figure 7 shows example results from further exploring this conclusion for non-stationary environments. Here the entries in the look-up tables for the rewiring of *B* connections in any dynamic nodes were re-assigned arbitrarily in offspring, i.e., not inherited from the parent. Fitness is significantly decreased (T-test, $p<0.05$) when $B<3$ and $K<4$. This again reinforces the view that evolution is able to exploit rewiring for non-chaotic networks to shape the attractor space, with the caveat it is most beneficial whilst the underlying fitness landscape is largely correlated, as perhaps might be expected (see Section 3.2); when the positions of fitness optima are structured, the rewiring is not simply a source of purely random variation which alters GRN behaviour such that they are better suited to non-stationary environments. Note also that the rewiring is still selected for in the $K>3$ cases indicating that rewiring can also serve as a useful source of (near) random variation in less structured fitness landscapes as well.

4.2 Inherited Rewiring: A Source of Evolutionary Innovation

In the above an offspring's nodes were initialized according to their genome specification as in the previous work on evolving *RBN*, regardless of the final connectivity pattern of the parent due to the effects of any dynamic nodes it contained. That is, genomic rearrangements were not inherited. Not least because of the considerable efforts invested by cells to avoid copy errors, it has long been argued that transposon-mediated changes are a principle source of heritable variation [Shapiro, 1992]. This can be explored with the model.

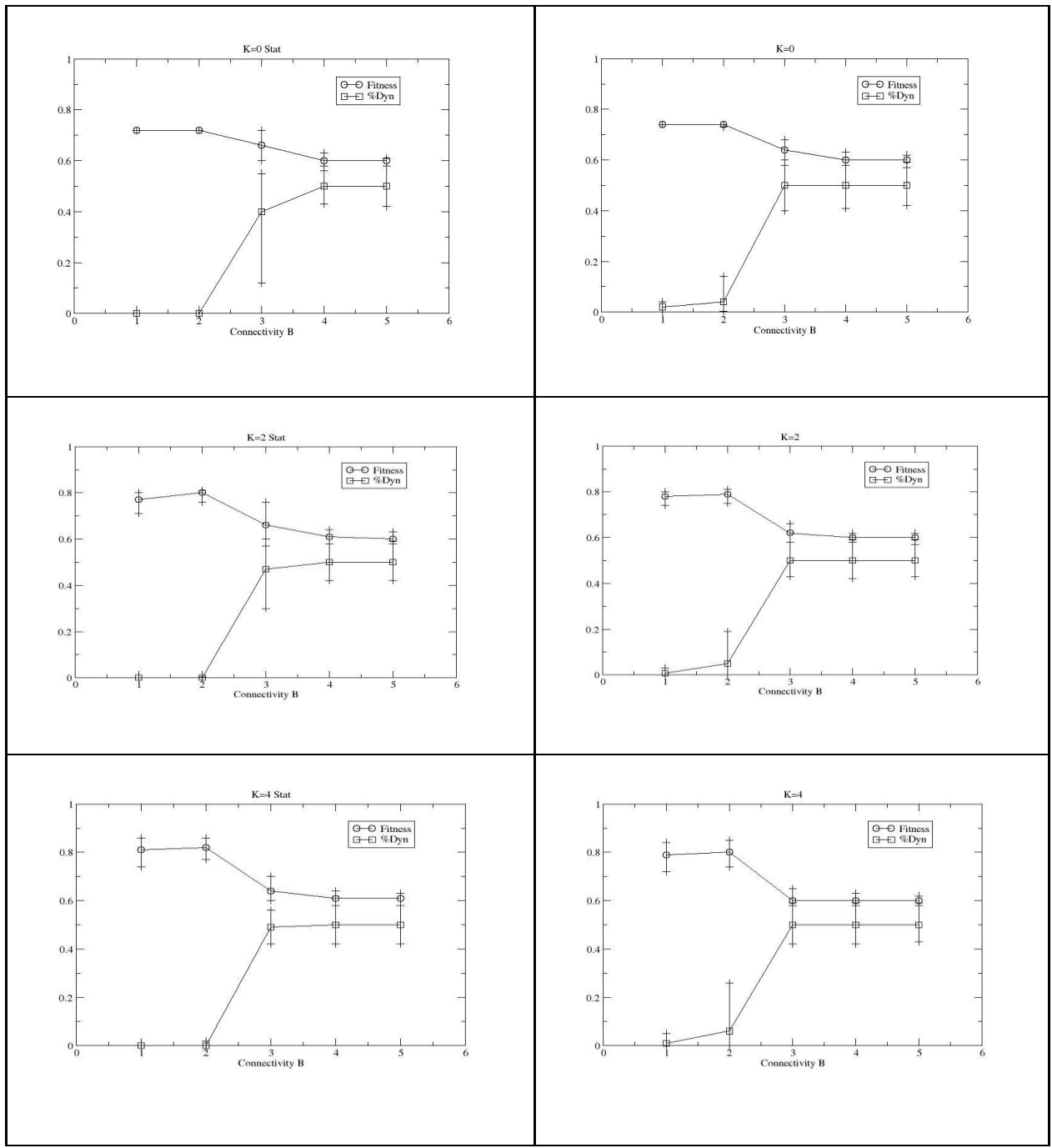

Figure 6: Example performance of the structurally dynamic *RBN* in stationary (left) and non-stationary (right) environments after 50,000 generations. Nodes capable of rewiring are selected for in the latter case.

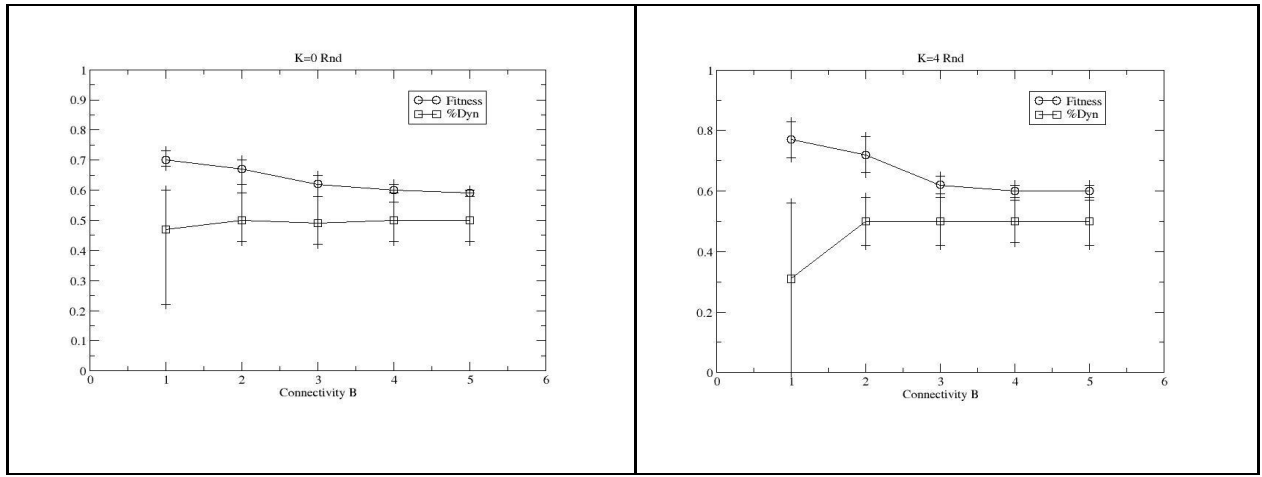

Figure 7: Example performance of the dynamic *RBN* where offspring do not inherit rewiring information.

Figure 8 shows examples of the evolutionary behaviour of the dynamic *RBN* when the parent's final network structure and node states are inherited by the offspring in the non-stationary case. The very first *RBN* are assigned random connectivity and node start states. The traditional random mutation operations were also reduced such that they can either: alter the Boolean function of a randomly chosen node; turn a node into or out of being a dynamic rewiring node; alter one of the rewiring entries in the look-up table if it is a dynamic node; or, alter a randomly chosen *B'* connection, again only if it is a dynamic node. That is, the rewiring behaviour becomes the principle source of connectivity variation for evolution, as opposed to random variation. The results indicate there is no significant change in fitness or the percentage of dynamic nodes (T-test, $p<0.05$) to the previous version for all *B* and *K* combinations used (compare to Figure 6). That is, *the evolutionary process appears equally able to exploit the use of genomic rearrangements to provide topological innovation as with directly applied mutations.*

It can be noted that in all known prior uses of transposons within simulated evolution their role has only been considered at the point of reproduction – either as a form of recombination (e.g., [Simoes & Costa, 1999]) or DNA sequence duplication (e.g., [Ferreira, 2001]) – and *not* as an on-going, context dependent restructuring processes during the lifecycle of the parent.

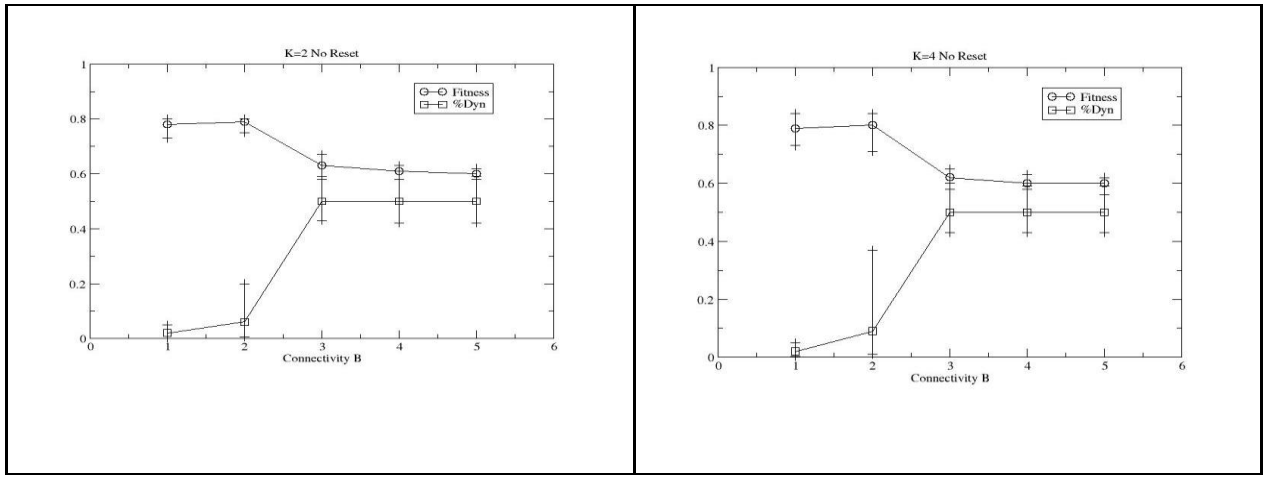

Figure 8: Example performance of the structurally dynamic *RBN* in non-stationary environments after 50,000 generations where offspring inherit structural changes made during the lifecycle of the parent.

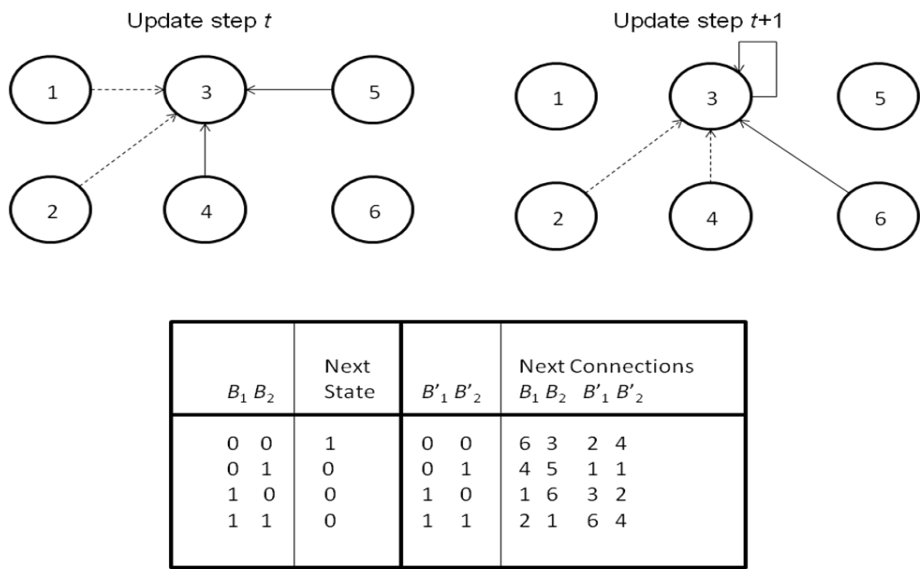

Figure 9: Example *RBN* with full structural dynamism. The look-up table and connections for node 3 are shown. Nodes capable of rewiring have *B'* extra structure regulation connections into the network (dashed arrows) and use the state of those nodes to alter the standard *B* transcription regulation connections (solid arrows) *as well as* their own connections on the next update cycle ($B'=2$). Thus in the *RBN* shown, at update step *t*, node 3 is shown using the states of nodes 4 and 5 to determine its state for the next cycle, and 1 and 2 for structural changes. Assuming both 4 and 5 are at state '0', the given node above would transit to state '1' for the next cycle. It would also source its *B* inputs from nodes 6 and 3 on that subsequent cycle, and source its *B'* inputs from nodes 2 and 4, all as defined in the first row of the table shown.

## 4.3 Full Structural Dynamism

In all of the above models, the node connections for structural regulation (*B'*) were fixed during the GRN lifecycle. That is, it was assumed the actions of genomic rearrangement could not disturb this aspect of the GRN. As shown in Figure 9, the topology of the structural regulation network component can also be allowed to rearrange in the same way as the transcription regulation component, i.e., such that both *B* and *B'* connections are rewired according to a node's table. In this way the traditional mutation can be further reduced so it can either: alter the Boolean function of a randomly chosen node; turn a node into or out of being a dynamic rewiring node; or, alter one of the rewiring entries in the look-up table if it is a dynamic node.

As the examples in Figure 10 indicate, the results again show there is no significant change in fitness (T-test, $p<0.05$) to the previous versions for all *B* and *K* combinations used in the non-stationary case. As Figure 10 also shows, examination of the percentage of dynamic nodes for *B*<3 indicates at least a doubling on average but there is wide variance in behaviour and it is therefore typically not a consistently statistically significant increase. However, *the trend for an increased use of structural dynamism with a decrease in the types of random mutation is clear*.

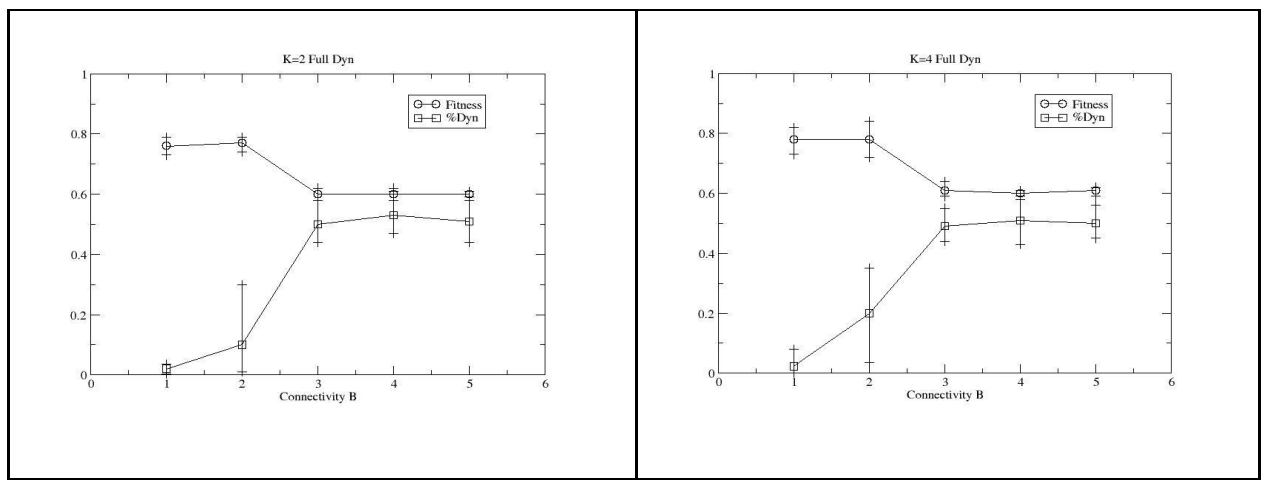

Figure 10: Example performance of the structurally dynamic *RBN* in non-stationary environments after 50,000 generations where offspring inherit structural changes made during the lifecycle of the parent, and where all aspects of connectivity are potentially varied.

## 5. Conclusions

For over 60 years, molecular biologists have been identifying a wide variety of mechanisms through which changes in DNA sequences occur in ways other than copy errors. Specific biochemical processes generate novelty through targeted DNA restructuring based upon the internal and external state of a GRN during the organismal lifecycle – what has been termed natural genetic engineering [Shapiro, 1992]. This paper has presented an extended version of a well-known abstract GRN model to begin to explore such mechanisms, both with respect to their temporal behaviour and potential to act as a source of evolutionary innovation.

It has been shown that rewiring can increases the average size of attractors for low connectivity *RBN* but not typically such that they become chaotic. When simulated evolution is used to design such GRN, it is found that structural dynamism is positively selected for in non-stationary environments but not simple stationary environments. Moreover, any genomic rearrangements occurring during a parent's lifecycle can be inherited by the offspring without detriment. This was shown to be the case even though the traditional source of direct (random) connection mutations was *removed*. The model still currently relies upon random changes to alter the structure regulatory circuits (table entries) and hence structure is still varied randomly indirectly through its actions. However, results here indicate that the structural dynamism is not acting in a purely random way since removing the inheritance of the rewiring table entries resulted in reduced fitness. It can also be noted that a rewiring process has also been explored which adjusts connections relative to their current position, e.g., within a range of +/- 5 nodes, as opposed to the explicit node changes used here with the same general behaviour seen (not shown).

Based on these findings, a number of extensions can be envisaged in the near future including the use of retrotransposon-like copying, as opposed to the DNA trasnsposon-like cutting and pasting here, thereby causing changes in the size of the GRN (see [Bull, 2012]), and consideration of altering the binding process, i.e., the Boolean function, at each node.